\documentstyle[preprint,prl,aps]{revtex}

\def\beq{\begin{equation}}
\def\eeq{\end{equation}}
\def\beqa{\begin{eqnarray}}
\def\eeqa{\end{eqnarray}}

\def\lsim{\mathrel{\raise.3ex\hbox{$<$\kern-.75em\lower1ex\hbox{$\sim$}}} }
\def\gsim{\mathrel{\raise.3ex\hbox{$>$\kern-.75em\lower1ex\hbox{$\sim$}}} }

\begin{document}
\draft

\hbadness=10000
\pagenumbering{arabic}

\preprint{{\vbox{\hbox{KEK-TH-642, 
NCKU-HEP-00-01, KIAS-P00014,
DPNU-00-13}}}} 
\vspace{1.0cm}

\title{\bf Fat penguins and imaginary penguins \\
in perturbative QCD}
\author{Yong-Yeon Keum$^{1}$
\footnote{Email: keum@phys.sinica.edu.tw}, Hsiang-nan Li$^{2}$
\footnote{Email: hnli@mail.ncku.edu.tw} and 
A.I. Sanda$^3$\footnote{Email: sanda@eken.phys.nagoya-u.ac.jp}
} 
\address{ $^{1}$ Institute of Physics, 
Academia Sinica, \\
 Taipei, Taiwan 115, Republic of China }
\address{ $^{2}$ Department of Physics, National Cheng-Kung University,\\
Tainan, Taiwan 701, Republic of China}
\address{ $^{3}$ Department of Physics, Nagoya University, Nagoya 464-01,
Japan}
\date{\today}
\maketitle

\vspace{10mm}
\begin{abstract}
We evaluate $B\to K\pi$ decay amplitudes in perturbative QCD
picture. It is found that penguin contributions are dynamically enhanced
by nearly 50\% compared to those assumed in the factorization
approximation. It is also shown that annihilation diagrams are not
negligible, and give large strong phases. Our results
for branching ratios of $B\to K\pi$ decays for a
representative parameter set are consistent with data. 
\end{abstract}
\vskip1.0cm

\thispagestyle{empty}

\newpage

Factorization assumption (FA) for nonleptonic two-body $B$ and $D$
meson decays pioneered by Stech and his collaborators \cite{BSW} has
been extremely successful. It gives correct order of magnitude for
branching ratios of most two-body $B$ meson decays. Why does it work so
well ? Recent CLEO data of $B\to \pi\pi$ and $B\to K\pi$ branching ratios
\cite{YK,CLEO3} require not only order-of-magnitude predictions but
quantitative predictions for these decay modes. As asymmetric $B$
factories, which are eventually capable of producing almost $10^8$ $B$'s
per year, have started their operation, quantitative theoretical
understanding will allow us to extract CP phases hidden in the above
branching ratios. How can we go beyond FA ?

Let us see what QCD can say about these questions. The fact that 5 GeV
of energy is released and shared by two light mesons suggests that the
basic interaction in two-body $B$ meson decays is mainly
short-distance. In this letter we shall attempt to compute these decay
amplitudes using as much information from the underlying theory, QCD, as
possible. Our method is perturbative QCD (PQCD) factorization theorem,
which has been worked out by Li and his collaborators \cite{LY1,L1,CL,YL}
based on the formalism developed by Brodsky and Lapage \cite{BL} and by
Botts and Sterman \cite{BS}.

Consider the specific $\bar{B}^0\to K^-\pi^+$ decay amplitude shown in
Fig.~1, where the $b\to s\bar{u} u$ decay occurs. The pair of the
$s$ and $\bar{u}$ quarks fly away and form the $K^-$ meson. The spectator
$\bar{d}$ quark of the ${\bar B}^0$ meson is more or less at rest and the
$u$ quark is flying away. The probability that a quark and an antiquark
with large relative velocity form the $\pi^+$ meson is suppressed by the
pion wave function. How big is the suppression ? It depends on the
functional form of the wave function. It is safe to say that this
suppression from the wave function is of the form
$\left({\Lambda_{\rm QCD}}/{M_B}\right)^n$, where $n$ is likely to be
large. Therefore, we expect that dominant contributions to the
$\bar{B}^0\to K^-\pi^+$ decay come from the process, where a
hard gluon is exchanged so that $\bar{d}$ quark momentum and $u$ 
quark momentum are aligned to form the pion. The rectangular dotted boxes
in Fig.~1 enclose the part of interaction which is hard. The blobs
represent wave functions giving amplitudes for a quark and an antiquark
to form a meson.

For the $B$ meson mass $M_B\gg \Lambda_{\rm QCD}$ and the kaon and
pion masses $M_K\sim M_\pi\sim 0$, the $\bar{B}^0\to K^-\pi^+$ decay
amplitude is then written as a convolution of four factors,
\begin{equation}
M=\int d[x]d[{\bf b}]\phi_B(x_1,{\bf b}_1)
\phi_K(x_2,{\bf b}_2)\phi_\pi(x_3,{\bf b}_3)
H([x],[{\bf b}],M_B)\;,
\label{2}
\end{equation}
where the wave functions $\phi_{B,K,\pi}$ for the
${\bar B}^0,~K^-,~\pi^+$ mesons absorb nonperturbative
dynamics of the process, the hard amplitude
$H([x],[{\bf b}],M_B)$ can be calculated in perturbation
theory, $[x]$ is a shorthand for the momentum fractions $x_1$, $x_2$, and
$x_3$ associated with the ${\bar d}$ quark in the $\bar{B}^0$ meson, the
$u$ quark, and the ${\bar d}$ quark in the $\pi^+$ meson, respectively,
and $[{\bf b}]$ is a shorthand for the two-dimensional vectors,
{\it i.e.}, the transverse extents, ${\bf b}_1$, ${\bf b}_2$, and
${\bf b}_3$ of the $\bar{B}^0$, $K^-$ and $\pi^+$ mesons, respectively.

Here are some important questions:
\begin{enumerate}
\item
Is $H([x],[{\bf b}],M_B)$ really dominated by short-distance
contributions ?
\item
Figure 1(a) is factorizable, since it can be written in terms of the
$B\to\pi$ transition form factor $F^{B\pi}$ and the kaon decay constant
$f_K$. This is the amplitude considered in FA. FA assumes that a
nonfactorizable amplitude from Fig.~1(b), which can not be written in
terms of a form factor and a decay constant, is negligible compared to
Fig.~1(a).

Are nonfactorizable amplitudes negligible compared to factorizable
ones ? We have made a numerical study of this issue. While it is
important to always check the relative magnitudes, we have found that
nonfactorizable contributions are usually less than few percents of
factorizable contributions. This is the reason FA has been so successful.
However, there are exceptions: 
(1) In $B\to D\pi$ decays some nonfactorizable contributions can reach as
much as 30\% of factorizable ones. Actually, the experimental fact
that the ratio $a_2/a_1\sim 0.2$ in the Bauer-Stech-Wirbel model
\cite{BSW} requires large nonfactorizable contributions. 
(2) In $B\to J/\psi K^{(*)}$ decays nonfactorizable and factorizable
contributions are of the same order of magnitude \cite{YL}.
\item
It is well known \cite{bigi} that the role of penguins is essential for
explaining the observed $B\to K\pi,~\pi\pi$ branching ratios. How big are
penguin amplitudes ? We shall show below that penguin amplitudes can be
dynamically enhanced by 50\% in PQCD compared to those assumed in FA.
\item
Are annihilation diagrams in Fig.~2 really negligible ?
\item
How big are final-state-interaction (FSI) effects ? It is impossible to
compute FSI phases in FA. Effects from infinite soft gluon exchanges
among mesons in two-body $B$ meosn decays have been analyzed
quantitatively by means of renormalization-group methods and found to be
small \cite{LT}. This observation implies that effects from exchange of
soft objects between the two final-state mesons are also small. Where
then do strong phases come from ? We shall show that contrary to common
belief, annihilation diagrams are important, and in fact, they contribute
large strong phases.

\end{enumerate}

We present the factorizable PQCD amplitudes $F_e$, $F_{e4}^P$, and
$F_{e6}^P$ corresponding to Fig.1(a) and $F_a$, $F_{a4}^P$, and
$F_{a6}^P$ corresponding to Fig.2(a) from the four-quark
operators $O_{1,2}$, $O_{3,4,9,10}$, and $O_{5,6,7,8}$, respectively
\cite{KLS},
\beqa
F^P_{e4}&=&16\pi C_FM_B^2\int_0^1 dx_1dx_3\int_0^{\infty} b_1d b_1b_3db_3
\phi_B(x_1,b_1)
\nonumber \\
& &\times\{\left[(1+x_3)\phi_\pi(x_3)+r_\pi(1-2x_3)\phi'_\pi(x_3)
\right]E_{e4}(t^{(1)}_e)h_e(x_1,x_3,b_1,b_3,M_B)
\nonumber\\
& &+2r_\pi\phi'_\pi(x_3)E_{e4}(t^{(2)}_e)
h_e(x_3,x_1,b_3,b_1,M_B)\}\;,
\label{int4}\\
F^P_{e6}&=&32\pi C_FM_B^2\int_0^1 dx_1dx_3\int_0^{\infty}b_1db_1b_3db_3
\phi_B(x_1,b_1)
\nonumber \\
& &\times r_K\{\left[\phi_\pi(x_3)+r_\pi(2+x_3)\phi'_\pi(x_3)
\right]E_{e6}(t^{(1)}_e)h_e(x_1,x_3,b_1,b_3,M_B)
\nonumber\\
& &+\left[x_1\phi_\pi(x_3)+2r_\pi(1-x_1)\phi'_\pi(x_3)\right]
E_{e6}(t^{(2)}_e)h_e(x_3,x_1,b_3,b_1,M_B)\}\;,
\label{int6}\\
F^P_{a4}&=&16\pi C_FM_B^2\int_0^1 dx_2dx_3\int_0^{\infty}b_2db_2b_3db_3
\nonumber \\
& &\times\{\left[-x_3\phi_{K}(x_2)\phi_\pi(x_3)
-2r_\pi r_K(1+x_3)\phi'_{K}(x_2)\phi'_\pi(x_3)\right]
E_{a4}(t^{(1)}_a)h_a(x_2,x_3,b_2,b_3,M_B)
\nonumber\\
& &+\left[x_2\phi_{K}(x_2)\phi_\pi(x_3)
+2r_\pi r_K(1+x_2)\phi'_{K}(x_2)\phi'_\pi(x_3)\right]
E_{a4}(t^{(2)}_a)h_a(x_3,x_2,b_3,b_2,M_B)\}\;,
\label{exc4}\\
F^P_{a6}&=&32\pi C_FM_B^2\int_0^1 dx_2dx_3\int_0^{\infty}b_2db_2b_3db_3
\nonumber \\
& &\times\{\left[r_\pi x_3\phi_{K}(x_2)\phi'_\pi(x_3)
+2r_K\phi'_{K}(x_2)\phi_\pi(x_3)\right]
E_{a6}(t^{(1)}_a)h_a(x_2,x_3,b_2,b_3,M_B)
\nonumber\\
& &+\left[2r_\pi\phi_{K}(x_2)\phi'_\pi(x_3)
+r_Kx_2\phi'_{K}(x_2)\phi_\pi(x_3)\right]
E_{a6}(t^{(2)}_a)h_a(x_3,x_2,b_3,b_2,M_B)\}\;,
\label{exc6}
\eeqa
$C_F$ being a color factor.
The expression of $F_e$ ($F_a$) for the $O_{1,2}$ contributions is the
same as $F^P_{e4}$ ($F^P_{a4}$) but with the Wilson coefficient
$a_1(t_e)$ ($a_1(t_a)$). The hard functions $h$'s in
Eqs~(\ref{int4})-(\ref{exc6}) are given by
\beqa
h_e(x_1,x_3,b_1,b_3,M_B)&=&K_{0}\left(\sqrt{x_1x_3}M_Bb_1\right)
\nonumber \\
& &\times \left[\theta(b_1-b_3)K_0\left(\sqrt{x_3}M_B
b_1\right)I_0\left(\sqrt{x_3}M_Bb_3\right)
+(b_1\leftrightarrow b_3)\right]\;,
\label{dh}\\
h_a(x_2,x_3,b_2,b_3,M_B)&=&\left(\frac{i\pi}{2}\right)^2
H_0^{(1)}\left(\sqrt{x_2x_3}M_Bb_2\right)
\nonumber \\
& &\times\left[\theta(b_2-b_3)
H_0^{(1)}\left(\sqrt{x_3}M_Bb_2\right)
J_0\left(\sqrt{x_3}M_Bb_3\right)
+(b_1\leftrightarrow b_3)\right]\;.
\label{ah}
\eeqa
The evolution factors
\beq
E_{ei}(t)=\alpha_s(t)a_i(t)\exp[-S_B(t)-S_\pi(t)]\;,\;\;\;\;
E_{ai}(t)=\alpha_s(t)a_i(t)\exp[-S_K(t)-S_\pi(t)]\;,
\eeq
arise from the summation of infinite infrared gluon emissions that
give double (Sudakov) logarithms and single logarithms connecting the
hard scales $t$ and the characteristic scales $1/b$ of the wave functions.
For the explicit expressions of the Sudakov exponents $S_B$, $S_K$, and
$S_\pi$, refer to \cite{LY1}. The hard scales $t$ are chosen as the
virtualities of internal particles in hard $b$ quark decay amplitudes,
\beqa
t^{(1)}_e={\rm max}(\sqrt{x_3}M_B,1/b_1,1/b_3)\;,& &\;\;\;\;
t^{(2)}_e={\rm max}(\sqrt{x_1}M_B,1/b_1,1/b_3)\;,
\nonumber\\
t^{(1)}_a={\rm max}(\sqrt{x_3}M_B,1/b_2,1/b_3)\;,& &\;\;\;\;
t^{(2)}_a={\rm max}(\sqrt{x_2}M_B,1/b_2,1/b_3)\;.
\label{ht}
\eeqa
It has been shown that this choice minimizes higher-order corrections to
exclusive QCD processes \cite{MN}. Equation (\ref{ht}) is consistent
with the fact that the hard scales $t$ and the evolution effects
related to running of $t$ should be process-dependent. The Wilson
coefficients are
\beqa
a_1=C_2+\frac{C_1}{N_c}\;,\;\;\;\;
a_{4(6)}=C_{4(6)}+\frac{C_{3(5)}}{N_c}+\frac{3}{2}e_q
\left(C_{10(8)}+\frac{C_{9(7)}}{N_c}\right)\;,
\label{wilc}
\eeqa
for the tree and the $(V-A)(V\mp A)$ penguins, respectively, $N_c$ being
the number of colors. $F_{e(a)6}^P$ has a different integrand from
$F_{e(a)4}^{P}$, reflecting the different helicity structures.

The factors $r_\pi$ and $r_K$,
\begin{eqnarray}
r_\pi=\frac{m_{0\pi}}{M_B}\;,\;\;\;\;
m_{0\pi}=\frac{M_\pi^2}{m_u+m_d}\;;
\;\;\;\;
r_K=\frac{m_{0K}}{M_B}\;,\;\;\;\;
m_{0K}=\frac{M_K^2}{m_s+m_d}\;,
\end{eqnarray}
are associated with the normalizations of the pseudoscalar wave functions
$\phi'$, where $m_u$, $m_d$, and $m_s$ are the current quark masses of
the $u$, $d$ and $s$ quarks, respectively, and $M_\pi$ and $M_K$ the
pion and kaon masses, respectively. The pseudovector and pseudoscalar
pion wave functions $\phi_\pi$ and $\phi'_\pi$ are defined in terms of
matrix elements of nonlocal operators 
$\langle 0|{\bar d}\gamma^-\gamma_5 u|\pi\rangle$ and
$\langle 0|{\bar d}\gamma_5 u|\pi\rangle$, respectively.
The kaon wave functions $\phi_K$ and $\phi'_K$ possess similar
definitions.

We employ the following set of meson wave functions as an illustrative
example:
\begin{eqnarray}
\phi_B(x,b)&=&N_Bx^2(1-x)^2
\exp\left[-\frac{1}{2}\left(\frac{xM_B}{\omega_B}\right)^2
-\frac{\omega_B^2 b^2}{2}\right]\;,
\label{os}\\
\phi_\pi(x)&=&\frac{3}{\sqrt{2N_c}}f_\pi x(1-x)[1+
0.8(5(1-2x)^2-1)]\;,
\label{pwf}\\
\phi_{K}(x)&=&\frac{3}{\sqrt{2N_c}}f_{K}
x(1-x)[1+0.51(1-2x)+0.3(5(1-2x)^2-1)]\;,
\label{mn}\\
\phi'_\pi(x)&=&\frac{3}{\sqrt{2N_c}}f_\pi x(1-x)\;,
\;\;\;\;\;\;
\phi'_K(x)=\frac{3}{\sqrt{2N_c}}f_{K}x(1-x)\;,
\label{phik}
\end{eqnarray}
with the shape parameter $\omega_B=0.3$ GeV and the decay constants
$f_\pi=130$ MeV and $f_K=160$ MeV. The normalization constant $N_B$ is
related to the $B$ meson decay constant $f_B=190$ MeV via
$\int \phi_B(x,b=0)dx=f_B/(2\sqrt{2N_c})$. 
$\phi_K$ is derived from QCD sum rules \cite{PB2}.
All other meson wave functions and $f_B$ are determined from the data of
the $B\to D\pi$, $\pi\pi$ decays and of the pion form factor \cite{KLS}.
Note that we have included the intrinsic $b$ dependence for the heavy
meson wave function $\phi_B$ but not for the light meson wave functions
$\phi_\pi$ and $\phi_K$. It has been shown that the intrinsic $b$
dependence of the light meson wave functions, resulting in only 5\%
reduction of the predictions for the form factor $F^{B\pi}$,
is not important \cite{LY1}. We do not distinguish the pseudovector and
pseudoscalar components of the $B$ meson wave functions under the heavy
quark approximation. 

$\bullet$ {\bf Is PQCD legitimate ?} \hspace{2mm}
We show that PQCD allows us to compute two-body decay amplitudes by
examining where dominant
contributions to the form factor $F^{B\pi}$ come from.
Figure 3 displays the fractional contribution as a function of
$\alpha_s(t)/\pi$. It is observed that 97\% of the contribution arises
from the region with $\alpha_s(t)/\pi<0.3$. Therefore, our PQCD results
are well within the perturbative region. This analysis implies that
$H([x],[{\bf b}],M_B)$ is dominated by short-distance contributions,
contrary to the viewpoint of Beneke, Buchalla, Neubert, and Sachrajda
(BBNS) in \cite{BBNS}, and
that the physical picture we have given for what is happening in Fig.~1 
is indeed valid.

$\bullet$ {\bf Fat penguins in PQCD:} \hspace{2mm}
Let us have a
careful look at the matrix elements of the penguin operators. It is
noticed that unlike $C_2$, $C_4$ and $C_6$ have a steep $\mu$ dependence.
In FA, amplitudes depend on the matching scale.
Normally, it is taken to be $m_b/2$, $m_b$ being the $b$ quark mass,
but there is no theoretical basis
for this choice. One of the main advantages in PQCD is that it provides
a prescription for choices of the hard scale $t$: $t$ should be
chosen as the virtuality of internal particles in a hard amplitude
in order to decrease higher-order corrections. A good fraction of
contributions then come from $t< m_b/2$, and penguin contributions
are enhanced. Numerically, this enhancement is given by:
\beq
\frac{(F^P_{e6})_{\rm PQCD}}{(F^P_{e6})_{\rm FA}}=1.6\;,
\hspace{4mm}
\frac{(F^P_{e4})_{\rm PQCD}}{(F^P_{e4})_{\rm FA}}=1.4\;,
\hspace{4mm}
\frac{(F_e)_{\rm PQCD}}{(F_e)_{\rm FA}}=1.0\;,
\label{12}
\eeq
where $(F)_{FA}$ represent the form factors evaluated in PQCD but
with the Wilson coefficients $C(t)$ set to $C(M_B/2)$. Equation
(\ref{12}) shows that penguin contributions are dynamically fattened by
about 50\%, and that the tree amplitudes from $O_{1,2}$ remains
invariant. Other sources of penguin enhancement are referred to
\cite{KLS}.

The enhancement due to the increase of $C_6(t)$ with decreasing
$t$ makes us worry that the contribution from the small $t$ region
may be important. This will invalidate the perturbative expansion of
$H([x],[{\bf b}],M_B)$. As a check, we examine the fractional
contribution to $F_{e6}^P$ as a function of $\alpha_s(t)/\pi$. The
results, similar to Fig.~3, indicate that about 90\% (80\%) of the
contribution comes from the region with ${\alpha_s(t)}/{\pi}<0.3$ (0.2).
Therefore, exchanged gluons are still hard enough to guarantee the
applicability of PQCD.

We emphasize that the penguin enhancement is crucial for the simultaneous
explanation of the $B\to K\pi$, $\pi\pi$ data using a unitarity angle
$\phi_3\sim 90^o$ \cite{KLS,LUY}. It has been shown \cite{WS}
that a simultaneous understanding of the data
$R={\rm Br}(B^0\to K^\pm \pi^\mp)/{\rm Br}(B^\pm\to K^0 \pi^\mp)\sim 1.0$
and ${\rm Br}(B^0\to\pi^\pm\pi^\mp)\sim 4.3\times 10^{-6}$ is difficult
in FA . The former indeed leads to $\phi_3\sim 90^o$. However, 
the latter leads to $\phi_3\sim 130^o$, even if $m_0$ is streched to
$m_0\sim 4$ GeV corresponding to $m_d=2m_u=3$ MeV.

$\bullet$ {\bf Imaginary annihilation penguins:} \hspace{2mm}
There has been a widely spread folklore that the annihilation diagrams
give negligible contribution due to helicity suppression, 
just as in $\pi\to e\overline\nu$ decay. That is, 
a left-handed massless electron and a right-handed antineutrino
can not fly away back to back because of angular momentum conservation.
However, this argument does not apply to $F_{a6}^P$. A left-handed quark
and a left-handed antiquark, for which helicities are dictated by the
$O_6$ operator, can indeed fly away back to back \cite{chang}.
These behaviors have been reflected by Eqs.~(\ref{exc4}) and (\ref{exc6}):
Eq.~(\ref{exc4}) vanishes exactly, if the kaon and pion wave functions
are identical, while the two terms in Eq.~(\ref{exc6}) are constructive.
Numerical results in Table I show that the strong phase associated with
$F^P_{a6}$ is nearly $90^\circ$. The large absorptive part arises from
cuts on the intermediate state $(s\bar{d})$ in the decay
$\bar{B}^0\to s\bar{d} \to K^-\pi^+$ shown in Fig.~2. The intermediate
state $(s\bar{d})$ can be regarded as being highly inelastic, if expanded
in terms of hadron states.

On the issue of FSI, Suzuki has argued that the invariant mass of the
$s\bar{d}$ pair in Fig.~2 is of order
$(\Lambda_{\rm QCD}M_B)^{1/2}\sim 1.2$ GeV \cite{suzuki}. Hence, the
$B\to K\pi$ decays are located in the resonance region and their strong
phases are very complicated. We have computed the average hard scale
of the $B\to K\pi$ decays, which is about 1.4 GeV, in agreement with
the above estimate. Since the outging
$s\bar{d}$ pair should possess an invariant mass larger than 1.4 GeV,
the processes are in fact not so close to the resonance region.
We could interpret that the $B\to K\pi$ decays occur via a six-fermion
operator within space smaller than $(1/1.4)$ GeV$^{-1}$. Though they are
not completely short-distance, the fact that over 90\% of contributions
come from the $x$-$b$ phase space with $\alpha_s(t)/\pi< 0.3$ allows us
to estimate the decay amplitudes reliably. We believe that the
strong phases can be computed up to about 20\% uncertainties, which
result in 30\% errors in predictions for CP asymmetries.

$\bullet$ {Br$(B\to K\pi)$:} \hspace{2mm} 
We present PQCD results of various $B\to K\pi$ branching ratios in
Table II, which are well consistent with the CLEO data \cite{CLEO3}.
These results are meant to be an example for a representative parameter
set such as the wave functions in Eqs.~(\ref{os})-(\ref{phik}),
which are determined from the best fit to the data of the
$B\to D\pi$, $\pi\pi$ and of the pion form factor, and $m_{0\pi}=1.4$ GeV
and $m_{0K}=1.7$ GeV \cite{KLS}. When all
other two-body decay modes are considered, we shall present an exhaustive
study of the entire parameter space allowed by data uncertainties.

$\bullet$ {\bf Comparision with the BBNS approach:}\hspace{2mm}
Here we compare our approach to exclusive nonleptonic $B$ meson decays
with the BBNS approach \cite{BBNS}. The differences between the two
approaches are briefly summarized below. For more details, refer to
\cite{KL}.

As stated before, the PQCD theory for exclusive processes was first
formulated by Brodsky and Lepage \cite{BL}. This formalism has been
criticized by Isgur and Llewellyn-Smith \cite{IL}, since involved
perturbative evaluations, based on expansion in terms of a large coupling
constant in the end-point region of momentum fractions, are not reliable.
Li and Sterman \cite{LS} pointed out that Sudakov resummation of large
logarithms associated with parton transverse momenta, which was worked
out by Botts and Sterman \cite{BS}, causes suppression in the end-point
region. This suppression is strong enough to render PQCD analyses
self-consistent at energy scales of few GeV. The above approach was then
extended by Li and his collaborators to exclusive $B$ meson decays in the
heavy meson limit \cite{LY1,L1,CL,YL}, where Sudakov suppression, cutting
off the infrared singularity in heavy-to-light transition form factors
\cite{KL}, is even more important. Our calculation of two-body
$B$ meson decays has followed the formalism developed by the above
authors. Therefore, the $B$-to-light-meson transition form factors at
maximal recoil are calculable in PQCD. In the BBNS approach, because
Sudakov suppression is not considered, the transition form factors
are not calculable and must be treated as inputs.

Another crucial difference is that in the PQCD formalism annihilation
diagrams are of the same order as factorizable diagrams in powers of
$1/M_B$, which are both $O(1/(M_B\Lambda_{\rm QCD}))$. The BBNS approach
follows FA, in which it has been assumed that factorizable contributions,
being $O(1/\Lambda^2_{\rm QCD})$, are leading, and all other contributions
such as annihilation and nonfactorizable diagrams, being
$O(1/(M_B\Lambda_{\rm QCD}))$, are next-to-leading. 
Factorizable and nonfactorizable contributions are considered by BBNS,
but annihilation are not. The difference is again traced back to the
inclusion of parton transverse momenta and of the Sudakov form factor in
our calculation. With Sudakov suppression, gluon exchanged in factorizable
diagrams are as hard as those in annihilation diagrams. Because of parton
transverse momenta, the internal particles in hard amplitudes may go
onto the mass shell at nonvanishing momentum fractions \cite{KL}. As a
consequence, annihilation diagrams lead to large imaginary contributions,
whose magnitudes are comparable to factorizable ones, and to large CP
asymmetries in the $B\to K\pi$ decays.

We emphasize that annihilation diagrams are indeed subleading in the
PQCD formalism as $M_B\to \infty$. This can be easily observed from
the hard functions in Eqs.~(\ref{dh}) and (\ref{ah}). When $M_B$
increases, the $B$ meson wave funciton in Eq.~(\ref{os}) enhances
contributions to $h_e$ from smaller momentum fraction $x_1$ as expected.
However, annihilation amplitudes proportional to $h_a$, being independent
of $x_1$, are relatively insensitive to the variation of $M_B$. Hence,
factorizable contributions become dominant and annihilation contributions
are subleading in the $M_B\to\infty$ limit. For $M_B\sim 5$ GeV, our
calculaiton shows that these two types of contributions are comparable.

We have shown that PQCD allows us to compute matrix elements of various
four-quark operators. While FA gives reliable estimates for $O_{1,2}$,
since their Wilson coefficients are nearly constant in the hard scale
$t$, matrix elements of the penguin operators are another story. We have
observed that PQCD results are larger than FA results by about 50\% for
the penguin operators, because of the $t$ dependence of the Wilson
coefficients. With the penguin enhancement in PQCD, the CLEO data
of the $B\to K\pi$, $\pi\pi$ branching ratios can be understood in a
more self-consistent way. We have also pointed out that penguin
annihilation diagrams are not negligible as claimed in FA. In fact, they
contribute large strong phases, which are essential for predictions of
CP asymmetries.

\skip2.0cm
\acknowledgements
This work was supported in part by Grant-in Aid for Special
Project Research (Physics of CP Violation) and by Grant-in Aid for
Scientific Exchange from the Ministry of Education, Science and Culture 
of Japan. The works of H.N.L. and Y.Y.K. were supported by the National 
Science Council of R.O.C. under the Grant Nos. NSC-89-2112-M-006-004 and
NSC-89-2811-M-001-0053, respectively. We thank H.Y. Cheng,
M. Kobayashi, and members of our PQCD group
for helpful discussions.
\skip2.0cm


\newpage
\begin{table}[ht]
\begin{tabular}{ccc}
Amplitude  & Real part & Imaginary part\\
\tableline 
$F_e$& 71.60 & 0\\
$F_e^P$&-6.18& 0\\
$F_a^P$&0.30&2.58\\
\end{tabular}
\label{TABLE11.1}
\caption{Amplitudes for the ${\bar B}^0\to K^-\pi^+$ decay in units
of $10^{-2}$ with $F_e^P=F_{e4}^P+F_{e6}^P$ and $F_a^P=F_{a4}^P+F_{a6}^P$.}
\end{table} 
\vskip 2.0cm

\begin{table}[ht]
\begin{tabular}{ccc}
Branching ratio&PQCD prediction ($10^{-6}$) & CLEO data (average)
($10^{-6}$)\\
\tableline
${\rm Br}(B^+\to K^0\pi^+)$ & $21.72$ & $18.2^{+4.6}_{-4.0}\pm 1.6$\\
${\rm Br}(B^-\to \bar{K}^0\pi^-)$ & $21.25$ & $18.2^{+4.6}_{-4.0}\pm 1.6$\\
${\rm Br}(B^0\to K^+\pi^-)$ & $24.19$ & $17.2^{+2.5}_{-2.4}\pm 1.2$\\
${\rm Br}({\bar B}^0\to K^-\pi^+)$ & $16.84$ & $17.2^{+2.5}_{-2.4}\pm 1.2$\\
${\rm Br}(B^+\to K^+\pi^0)$ & $14.44$ & $11.6^{+3.0+1.4}_{-2.7-1.3}$\\
${\rm Br}(B^-\to K^-\pi^0)$ & $10.65$ & $11.6^{+3.0+1.4}_{-2.7-1.3}$\\
${\rm Br}(B^0\to K^0\pi^0)$ & $11.23$ & $14.6^{+5.9+2.4}_{-5.1-3.3}$\\
${\rm Br}({\bar B}^0\to \bar{K}^0\pi^0)$ & $11.84$ &
$14.6^{+5.9+2.4}_{-5.1-3.3}$
\end{tabular}
\caption{PQCD predictions of branching ratios for a representative
parameters set.}
\end{table}

\newpage


{\bf \Large Figure Captions}
\vspace{10mm}

\noindent
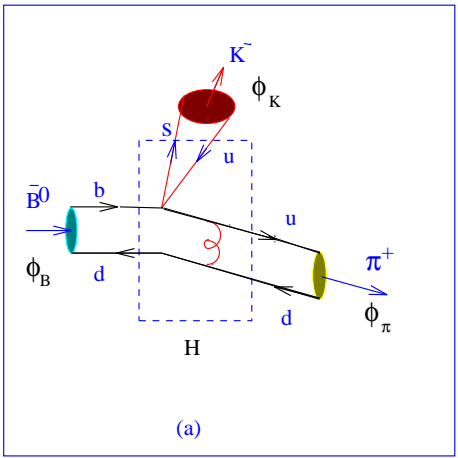
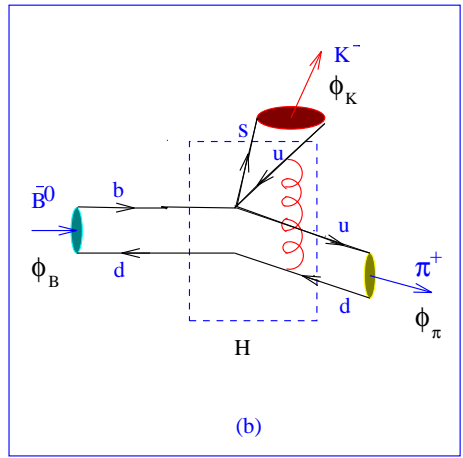
Fig. 1: (a) Factorizable and (b) nonfactorizable tree or penguin
contributions.
\vskip 0.5cm

\noindent
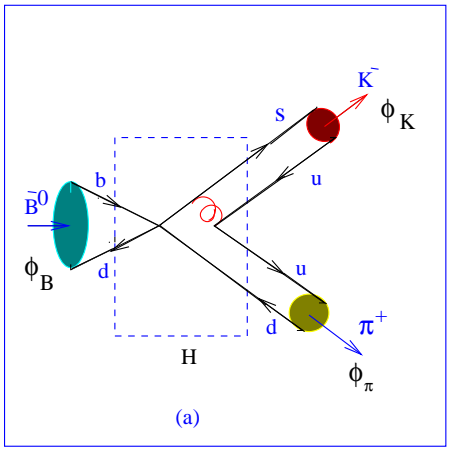
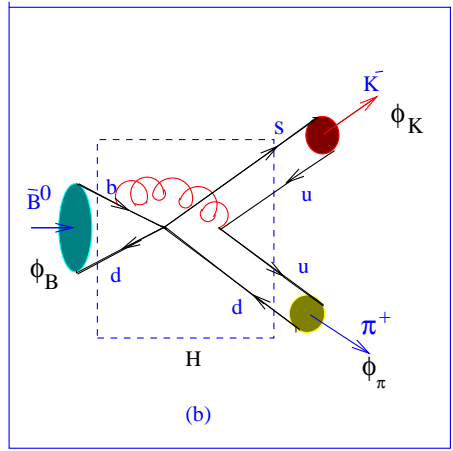
Fig. 2: (a) Factorizable and (b) nonfactorizable annihilation
contributions. A cut on the $s{\bar d}$ quark lines corresponds to
the absorptive part.
\vskip 0.5cm

\noindent
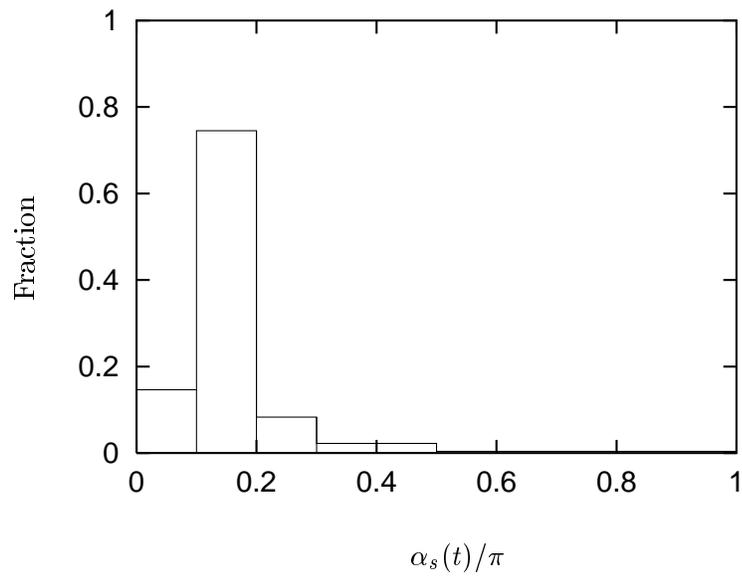
Fig. 3: Fractional contribution to the $B\to\pi$ transition form factor
$F^{B\pi}$ as a function of $\alpha_s(t)/\pi$.
\vskip 0.5cm



\end{document}